\documentclass[aip, amsmath,amssymb, reprint,twocolumn]{revtex4-1}

\usepackage{graphicx}
\usepackage{dcolumn}
\usepackage{bm}
\usepackage{epstopdf}
\usepackage{epsfig}
\usepackage{dcolumn}
\usepackage{float}
\usepackage{xcolor}

\begin{document}

\preprint{AIP/123-QED}

\title[Nm-sized cryogenic hydrogen clusters for a laser-driven proton source]{Nm-sized cryogenic hydrogen clusters for a laser-driven proton source}

\author{S. Grieser}
\email{s\_grie06@wwu.de}
\affiliation{Institut f{\"u}r Kernphysik, Westf{\"a}lische Wilhelms-Universit{\"a}t M{\"u}nster, 48149 M{\"u}nster, Germany}

\author{B. Aurand}
\email{bastian.aurand@hhu.de}
\affiliation{Institut f{\"u}r Laser- und Plasmaphysik, Heinrich-Heine-Universit{\"a}t, 40225 D{\"u}sseldorf, Germany}

\author{E. Aktan}
\affiliation{Institut f{\"u}r Laser- und Plasmaphysik, Heinrich-Heine-Universit{\"a}t, 40225 D{\"u}sseldorf, Germany}

\author{D. Bonaventura}
\affiliation{Institut f{\"u}r Kernphysik, Westf{\"a}lische Wilhelms-Universit{\"a}t M{\"u}nster, 48149 M{\"u}nster, Germany}

\author{M. B\"uscher}
\affiliation{Peter Gr{\"u}nberg Institut, PGI-6, Forschungszentrum J{\"u}lich, 52428 J{\"u}lich, Germany}

\author{M. Cerchez}
\affiliation{Institut f{\"u}r Laser- und Plasmaphysik, Heinrich-Heine-Universit{\"a}t, 40225 D{\"u}sseldorf, Germany}

\author{I. Engin}
\affiliation{Peter Gr{\"u}nberg Institut, PGI-6, Forschungszentrum J{\"u}lich, 52428 J{\"u}lich, Germany}

\author{L. Le{\ss}mann}
\affiliation{Institut f{\"u}r Kernphysik, Westf{\"a}lische Wilhelms-Universit{\"a}t M{\"u}nster, 48149 M{\"u}nster, Germany}

\author{C. Mannweiler}
\affiliation{Institut f{\"u}r Kernphysik, Westf{\"a}lische Wilhelms-Universit{\"a}t M{\"u}nster, 48149 M{\"u}nster, Germany}

\author{R. Prasad}
\affiliation{Institut f{\"u}r Laser- und Plasmaphysik, Heinrich-Heine-Universit{\"a}t, 40225 D{\"u}sseldorf, Germany}

\author{O. Willi}
\affiliation{Institut f{\"u}r Laser- und Plasmaphysik, Heinrich-Heine-Universit{\"a}t, 40225 D{\"u}sseldorf, Germany}

\author{A. Khoukaz}
\affiliation{Institut f{\"u}r Kernphysik, Westf{\"a}lische Wilhelms-Universit{\"a}t M{\"u}nster, 48149 M{\"u}nster, Germany}

\date{\today}

\begin{abstract}
A continuous cryogenic hydrogen cluster-jet target has been developed for laser-plasma interaction studies, in particular as a source for the acceleration of protons. Major advantages of the cluster-jet target are the compatibility with pulsed high repetition lasers and the absence of debris. The cluster-jet target was characterized using the Mie-scattering technique allowing to determine the cluster size and to compare it with an empirical prediction. In addition, an estimation of the cluster beam density was performed. The system was implemented at the high power laser system ARCTURUS and first successful measurements show the acceleration of protons after irradiation by high intensity laser pulses with a repetition rate of five Hertz.  
 
\end{abstract}

\maketitle

\section{Introduction}
In the field of laser-induced ion acceleration, stability and applicability have tremendously increased over the last two decades. Starting from the process of target-normal-sheath acceleration \cite{Wil01, Mor03} using flat metal foil targets, nowadays, a variety of different acceleration schemes have been identified. In the route of improving the parameters of laser-driven proton beams including flux, energy distribution, and collimation, new acceleration regimes have been proposed, like radiation-pressure-acceleration (RPA) \cite{Hen09a, Yan08, Aur13a} or break-out-afterburner (BOA) and more sophisticated targets have been proposed \cite{Aur13b, Aur14a}. A detailed review on the recent progress achieved in the area of laser-ion acceleration can be found in the publications of \textit{A. Macchi} \cite{Mac13a, Mac17}.\\
Two major disadvantages of most targets are the fact that they are made for single use, meaning they need to be replaced after each laser shot and that a large part of the incident laser energy is lost by the escape of energetic electrons transversely to the laser propagation direction \cite{Tre11, prencipe}.
Ideas to overcome the first problem are for example the use of a gas target, as described by \textit{M. Helle} \cite{Hel17}. The drawback is the low mass density resulting in a low particle flux. Other ideas are the use of a liquid or solid, dispersed in a jet-like structure \cite{Kar03, Schn05}. This guarantees an unlimited amount of target material at solid-state density but comes with the difficulty of proper laser alignment onto the often very fine structure. The second problem, the energy loss by escaping electrons, could be overcome by so-called mass-limited targets, e.g. using a very fine support structure or fully levitating targets \cite{Hil18}. Those targets face again the problem of replacement after each laser shot.\\
In contrast, a cluster-jet target overcomes most of the difficulties mentioned above. The cluster source delivers on the one hand mass-limited targets in a continuous beam, since every cluster serves as an isolated target. On the other hand, the cluster density is high enough to have a sufficient number of clusters in the focal volume during the interaction, which guarantees a high shot-to-shot stability. The clusters can be produced from a liquid \cite{Pra12} or cryogenic cooled gas \cite{Jin17}. 
A major benefit in the use of hydrogen gas is the high purity of the source allowing only protons to be accelerated in contrast to most other targets containing hydrocarbons where also heavier ions can be accelerated. \\
Recently, the use of micrometer-sized clusters of hydrogen for proton acceleration were demonstrated by \textit{Jinno et al.} \cite{Jin18}. Presented in this publication are the development and first successful measurements using a novel cluster-jet target for laser-driven proton acceleration, aiming for a reliable, high repetition rate with low maintenance for future applications. After implementation, the characterization of the target parameters such as cluster size using the technique of Mie-scattering has been performed. The performance of the target by irradiation with a laser beam at an intensity of $I$=10$^{20}$\,W/cm$^2$ has been successfully demonstrated and protons in the keV range were observed.

\section{The cluster-jet target}
\begin{figure}
\includegraphics[scale=0.42, angle = 90]{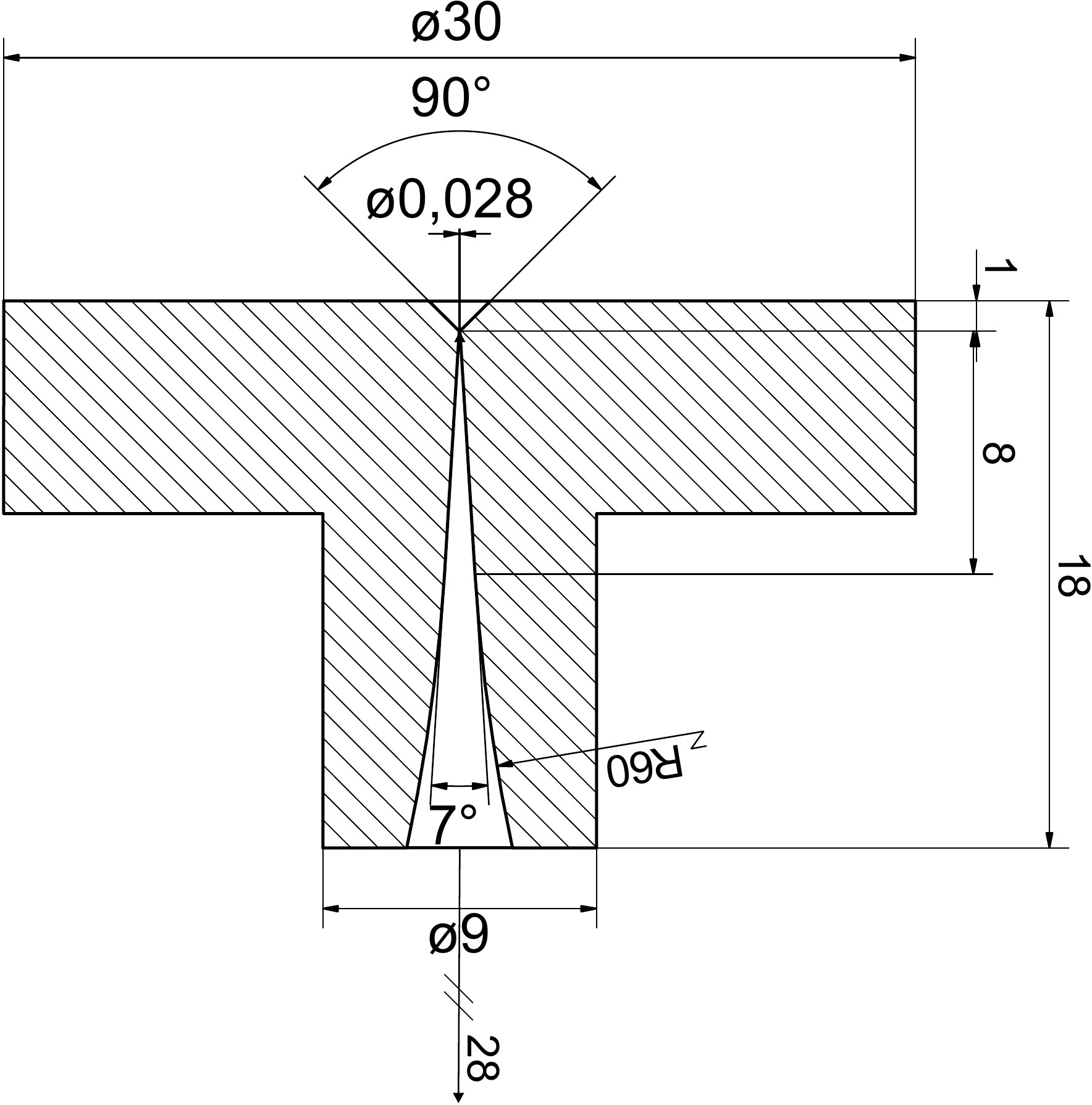}
\caption{\label{fig:Nozzle}Sketch of the used Laval nozzle. For hydrogen an inner diameter of $\o_{\text{H}_2} = 28 \, \mu \text{m}$ was used. The dimensions are given in units of mm.}
\end{figure}

The cluster-jet target MCT1S was designed and built by the University of M{\"u}nster to perform detailed studies on the production of cluster-jet beams. In a first set of measurements, a hydrogen cluster beam was well defined by a set of skimmers and studied of up to $2 \, \text{m}$ behind the target beam generator \cite{PhdTaschna}. Recently, the cluster-jet target MCT1S was revised and optimized in order to perform detailed investigations on the cluster-laser interaction very close to the target generator and at high target thicknesses, i.e. $>$ $10^{16} \, \text{atoms/cm}^3$ \cite{MAGrieser, PhDGrieser}. Depending on the experimental setup, this cluster-jet target offers the possibility to cool down gaseous target materials to temperatures of roughly $T= 20 \, \text{K}$. For the experiments presented here the target was modified to allow for interaction studies directly behind the nozzle, which increased the lowest temperature to roughly $T= 57 \, \text{K}$. The (pre-cooled) gas is directed with pressures of e.g. $p= 16 \, \text{bar}$ through a fine Laval nozzle with typical diameters of $28 \, \mu \text{m} - 100 \, \mu \text{m}$. With a starting temperature well below the inversion temperature and due to the specific convergent-divergent shape of the nozzle, the gas is further cooled down during passing the nozzle due to adiabatic cooling and the formation of clusters develops. \\
The Laval nozzle is one main component of a cluster-jet target and its complex inner geometry is essential for the cluster production process. The convergent short $90 \, ^\circ$ inlet cone merges after the narrowest point of the nozzle in a $17 \, \text{mm}$ long divergent outlet zone with first a constant opening angle followed by a radius (cf. Fig. \ref{fig:Nozzle}). The nozzles used here originate from a nozzle production line of CERN (Conseil europ\'{e}en pour la recherche nucl\'{e}aire). Based on this production process, an improved production process was developed at the University of M{\"u}nster to ensure the availability and the diversity of Laval nozzles for upcoming experiments using a cluster-jet target with different gases and stagnation conditions.\\

The cluster-jet target can be used with different gases as target material. Depending on the used target material, the nozzle geometry has to be adapted in order to provide appropriate target thicknesses and gas flows through the nozzle for the different gases. Thereby, the general geometry of the nozzle was the same and only the narrowest inner diameter has to been varied. In case of hydrogen, the nozzle had an inner diameter of $\o_{\text{H}_2} = 28 \, \mu \text{m}$.\\
In order to offer the possibility to cool down the target gas before entering the nozzle to temperatures of about $T= 60 \, \text{K}$, required to increase the target density as well as the cluster size, a two-stage cold head is used (cf. Table \ref{tab:devices}). The applied cold head works with a closed helium cycle in combination with a compressor. Both parts are connected via flexible high pressure lines. A displacer inside the cold head separates two volumes. According to the Gifford-McMahon process the movement of the displacer leads to a compression and expansion of the helium within the volumes resulting in a heat exchange with the compressor. The first stage of the cold head, the warm stage, pre-cools the gas and the second stage, the cold stage, realizes the final low temperatures of the nozzle. \\

\begin{table}
\caption{\label{tab:devices}Target components and their specification.}
\begin{ruledtabular}
\begin{tabular}{c|c}
target components & specification \\
 	 \hline
cold head & Leybold RGD $1245$\\	
compressor & Leybold Coolpak $6000$H \\
pressure controller  & SLA $5810$ (Brooks)\\
flow meter  & SLA $5860$ (Brooks)\\
hydrogen purifier  & HP-$50$ (Johnson Matthey) \\
baratron & $722$ (MKS)\\
temperature control & Lakeshore $206$/$336$\\
				 & PSP-$603$ (GwInstek) \\
temperature diodes & 1. Stage: DT 670 (Lakeshore)\\
					& 2. Stage: DT 670 (Lakeshore)\\
heating cartridges & 1. Stage: 50 W (Lakeshore)\\
					& 2. Stage: 100 W (Lakeshore)\\
 \end{tabular}
\end{ruledtabular}
\end{table}

\begin{figure}
\includegraphics[scale=0.32]{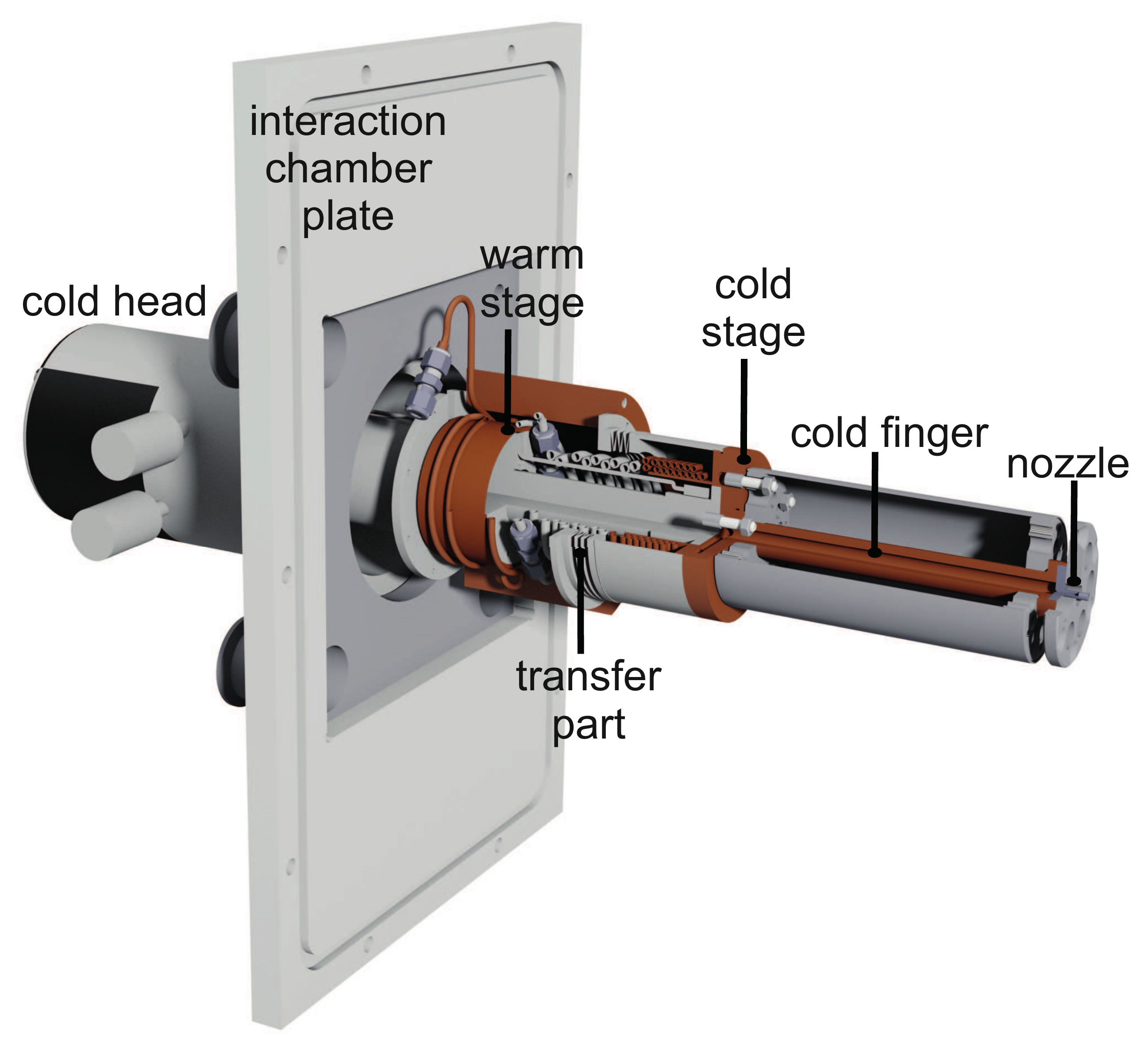}
\caption{\label{fig:MCT1S}The cluster-jet target MCT1S installed at a chamber plate of the interaction chamber. The motor of the cold head is located outside of the interaction chamber, while the main parts of the cluster-jet target including the cryogenic parts of the cold head extend into the chamber. A cold finger between the cold stage and the nozzle is installed to realize an interaction of the laser with the cluster beam directly behind the nozzle in the center of the interaction chamber. The cluster beam direction is from left to right.}
\end{figure}

\begin{figure}
\includegraphics{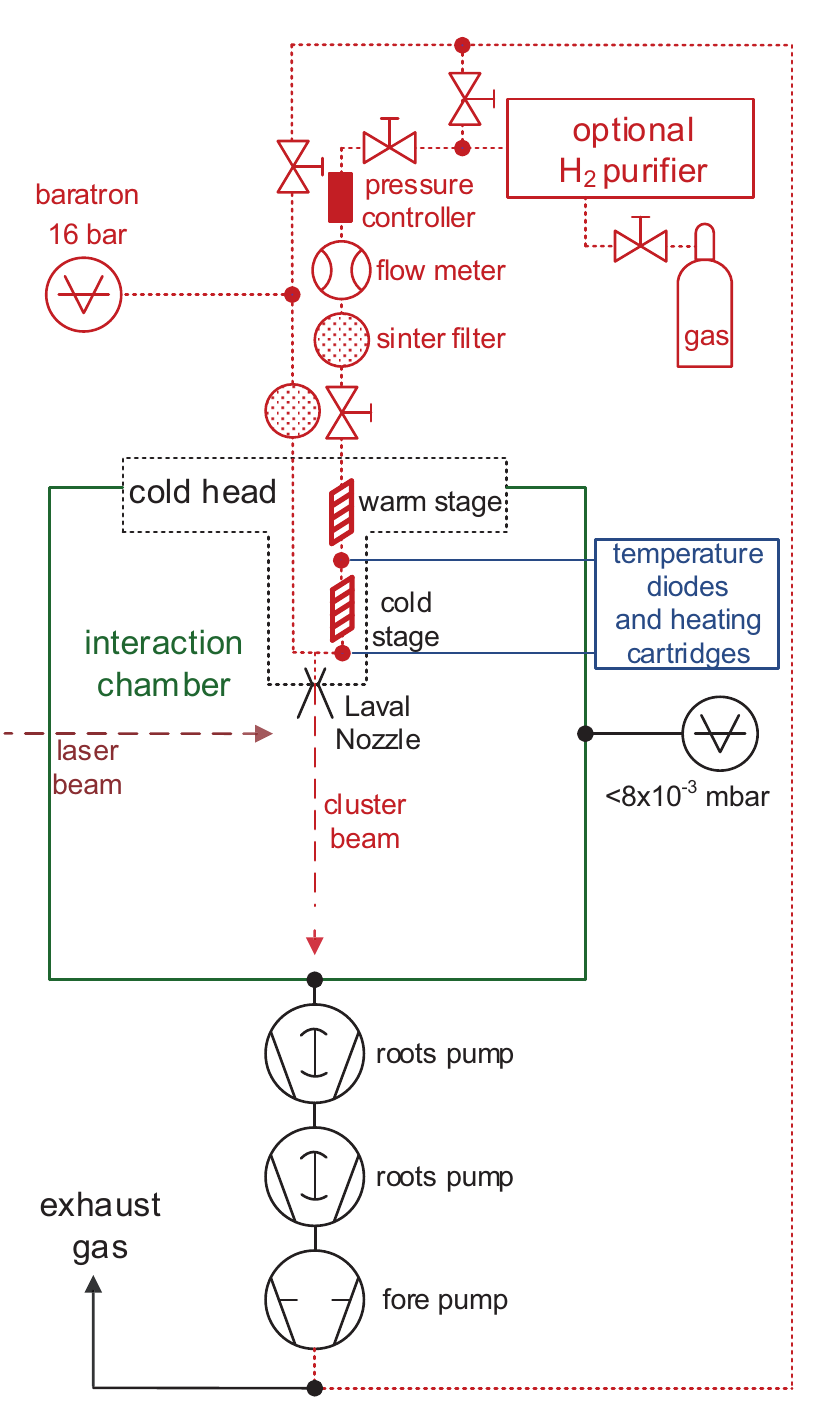}
\caption{\label{fig:Visio}Sketch of the gas (red) and vacuum system of the cluster-jet target MCT1S installed at the interaction chamber. A cold head is used to cool down the target gas. A pressure controller, a flow meter, temperature diodes and heating cartridges are installed to monitor and control the target parameters. The interaction zone between the laser and the cluster beam is located directly behind the Laval nozzle. Opposite to the Laval nozzle, a powerful roots pump is installed to pump the clusters directly after the interaction to ensure a minimal gas background.}
\end{figure}

Before entering the cold head, the target gas is guided through a pressure controller, to monitor and control the pressure of the gas in front of the nozzle, and is then fed to a flow meter (see Fig. \ref{fig:Visio}). The information of the flow meter is used to calculate the minimum diameter of the nozzle at a given pressure which offers the possibility to recognize a blocked nozzle. Due to the very small diameter of the nozzle of, e.g., $\o_{\text{H}_2} = 28 \, \mu \text{m}$ for hydrogen, the gas is additionally purified using a palladium hydrogen purifier. Therefore, the hydrogen gas has a purity of $9.0$ directly behind the device. The target gas is directed through pipes wrapped around the two stages of the cold head (see Fig. \ref{fig:MCT1S}). The pipes located at the cooling parts of the cold head are made out of copper to guarantee an optimal heat exchange between cold head and gas. The gas pipes in between the cooling parts are made of stainless steel to reduce a direct heat exchange between the cryogenic stages for an optimal cooling of the gas. Two additional sinter filters are installed to avoid a blocking of the nozzle due to possible impurities within the piping system of the target gas. Finally, the cooled gas enters a cold finger mounted directly on top of the cold stage of the cold head. The cold finger is used to place the nozzle exit and by this the interaction point between cluster and laser beam in the center of the interaction chamber. The cold finger consists of copper and is surrounded by a small walled hollow cylinder made out of reflecting stainless steel acting as heat radiation shielding. On the end of the cold finger, the Laval nozzle is mounted. Both, the cold finger as well as the nozzle are sealed with a circled indium sealing. This sealing is also gas tight at cryogenic temperatures and, therefore, ideally suited as nozzle sealing in the temperature range of importance here.\\

At the two stages of the cold head, temperature diodes and heating cartridges are installed to monitor and control the temperatures of both stages. The equipment of the cluster-jet target is remotely controlled by a slow control system realized in LabVIEW by National Instruments \cite{NI}, which is also used to display the target parameters on a website and to store them every $10 \, \text{s}$. With this system, it could be shown, that the stagnation conditions and consequently the characteristics of the cluster beam achieve a stability of better than $ \Delta T= 0.1 \, \text{K}$ and  $ \Delta p= 0.1 \, \text{bar}$. \\

\begin{figure*}
\includegraphics[scale=0.7]{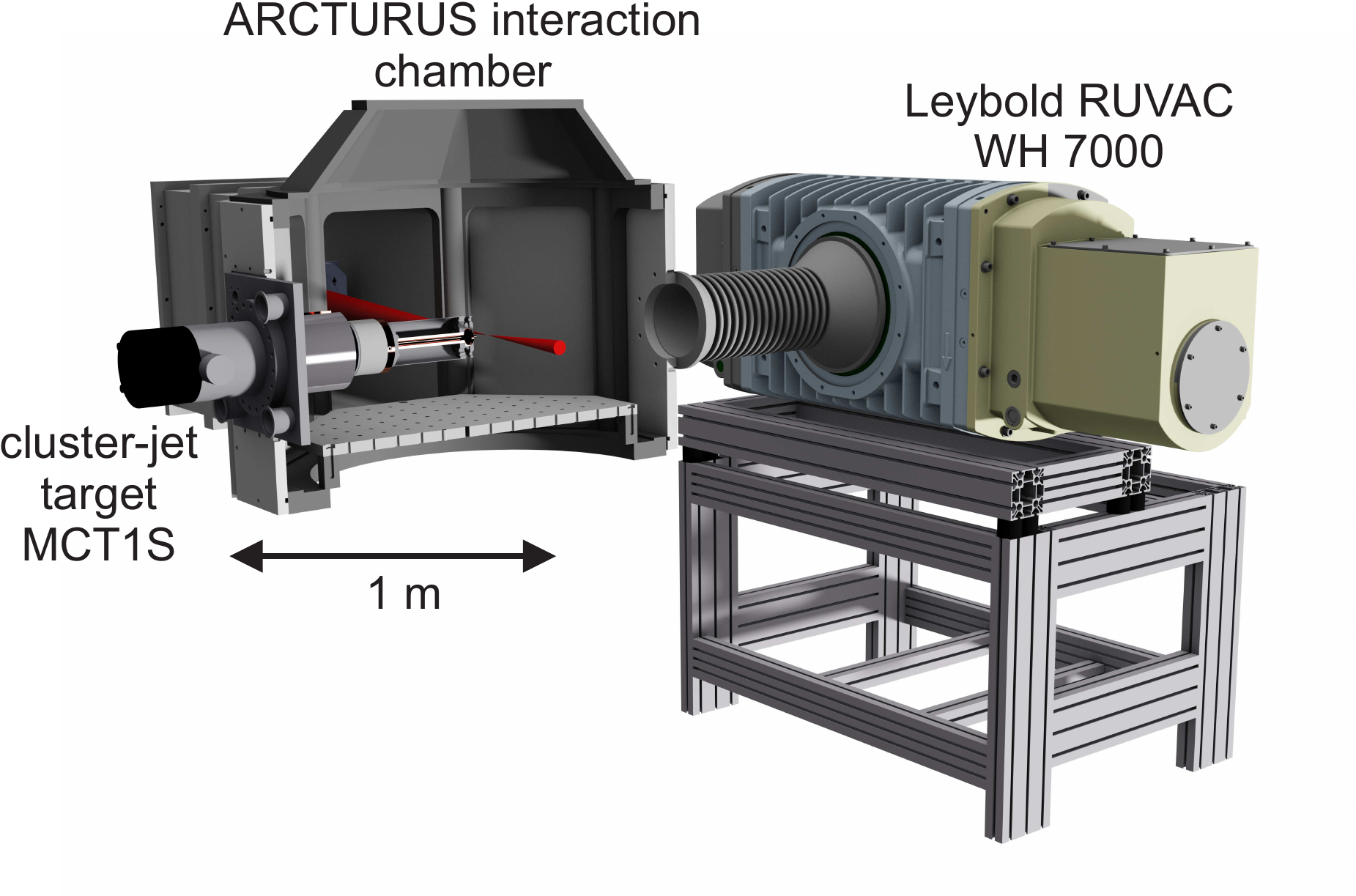}
\caption{\label{fig:MCT1SArcturus}The cluster-jet target MCT1S (left) installed at the interaction chamber of the ARCTURUS laser system in D{\"u}sseldorf. The cluster beam direction is from left to right and the interaction with the laser (red) is placed in the center of the chamber directly behind the cluster nozzle. After the interaction, the cluster beam enters a roots pump of type Leybold RUVAC WH $7000$ and gets pumped away.}
\end{figure*}

For the experiments presented here, the cluster-jet target is directly integrated in a removable interaction vacuum chamber plate and extends appropriately into the interaction chamber. For this reason, the laser pulse can be directly focused onto the cluster beam with no disturbing material like foils or walls in between. By this, the full laser performance is available and no disturbances occur. Moreover, if necessary, this setup allows for a fast and simple disassembly and installation of the cluster-jet target at the interaction chamber. Furthermore, even after removing and re-installation of the target, the position of the nozzle shows up exactly at the same position and no further re-adjustment of the laser focus with respect to the cluster beam is required. \\
The setup presented here allows for a very quick and simple change between different gases as target material. Depending on the gas, at maximum only the nozzle has to be exchanged and this can be done without dismounting the complete target. As a result, a series of measurements with various gas types can be performed rapidly one after the other.\\  
Opposite to the cluster-jet target, a powerful roots pump is installed to directly pump away the cluster beam after the interaction with the laser beam to ensure a good vacuum in the interaction chamber (Fig. \ref{fig:MCT1SArcturus}, Table \ref{tab:pumps}). With this powerful pumping system the vacuum in the interaction chamber was below $ 8 \times 10^{-3} \, \text{mbar}$ during the performed measurements at hydrogen gas flows of $ 0.016 \, \text{l}_\text{n} / \text{s}$.\\

\begin{table}
\caption{\label{tab:pumps}Vacuum pumps used at the cluster-laser interaction chamber and their pumping speed.}
\begin{ruledtabular}
\begin{tabular}{c|c}
vacuum pump & nominal pumping speed \\
 	 \hline
Leybold SOGEVAC SV $300$ & $ 240 \, \text{m}^3/ \text{h}$\\	
Leybold RUVAC WSU $1001$ & $ 1000 \, \text{m}^3/ \text{h}$\\
Leybold RUVAC WH $7000$ & $ 7800 \, \text{m}^3/ \text{h}$\\
\end{tabular}
\end{ruledtabular}
\end{table}

\section{Cluster-jet characterization}
In order to investigate the laser-cluster interaction and the underlying physical processes, it is necessary to determine the cluster beam characteristics. Of special interest is the target volume density at the position of the laser interaction point. Although a direct measurement in principle is possible, due to the experimental setup such a measurement would be accompanied here with large systematic uncertainties. Alternatively, a much more reliable method is an estimation using the gas flow through the nozzle. Since the gas flow meter information was designed for higher gas flows at lower temperatures the gas flow is calculated based on the well known stagnation conditions and the nozzle geometry. The gas volume flow $q_v$ through the nozzle can be calculated using \cite{Wutz1988}

\begin{equation}
q_v = A^* \frac{p_0}{\sqrt{M T_0}} \frac{T_\text{N}}{p_\text{N}} \left( \frac{2}{\kappa + 1} \right)^{\frac{\kappa +1}{2 \left( \kappa -1 \right)}} \sqrt{\kappa R}\,.
\label{flow}
\end{equation}

Therein, $A^*$ is the cross section of the narrowest point of the nozzle, $p_0 $ and $T_0 $ are the stagnation conditions of the gas in front of the nozzle,  $p_\text{N}= 1.01325  \, \text{bar}$ and $T_\text{N} = 273.15  \, \text{K}$ are the normal pressure and temperature, $M$ is the molar mass of the gas, $\kappa = C_p / C_V $ is the heat capacity ratio and $R$ is the universal gas constant. A calculation of the mass flow $\dot m$ with respect to the gas flow $q_v$

\begin{equation}
\dot m=\frac{q_v \cdot M \cdot p_\text{N}}{R \cdot T_\text{N}}
\end{equation}

allows for a determination of the target volume density $\rho_\text{volume}(x)$ with the cluster beam cross section $A_\text{beam}\left( x \right)$ at the distance $x$ to the nozzle and the Avogadro constant $N_{\text{A}}$
	
\begin{equation}
\rho_\text{volume}(x)=\frac{\dot m}{v \cdot A_\text{beam}\left( x \right)}\cdot \frac{N_{\text{A}}}{M}\,.
\label{thickness}
\end{equation}
	
For the temperatures used, in good approximation the mean cluster velocity $v$ can be estimated assuming an ideal gas \cite{PhdTaschna}
	
\begin{equation}
v=\sqrt{\frac{2 \kappa}{\kappa -1} \frac{RT_0}{M}}\,.
\end{equation}

In the case of a real gas the jet velocity might be lower by, e.g., a few percent. Therefore, the estimated target density might be larger by this number. Obviously the target density depends on the used gas and the stagnation conditions, which can easily be defined just by changing the nozzle temperature and/or the gas pressure. In good approximation the propagation of the cluster beam after leaving the nozzle is based on the intercept theorem \cite{PhdKoehler}. Assuming a circular volume element of the cluster beam increases with increasing distance from the nozzle, results in a density decreases by $ \sim 1/x^2$ with increasing distance $x$. Taking this into account, the presented measurements were performed directly behind the nozzle at $28 \, \text{mm}$ with respect to the nozzle exit. The estimated target densities for the presented measurements at the interaction point are listed in Table \ref{tab:Cluster}. The cluster beam diameter at the interaction point was calculated considering the opening angle of the nozzle of $\alpha =  7 \, ^\circ$ as cluster beam divergence. However, it has to be noted that previous measurements at much lower temperatures of, e.g., $25 \, \text{K}$ and with liquid hydrogen entering the nozzle allowed for an optical observation of the jet beam \cite{PhdKoehler}. These studies resulted in a smaller divergence in the range of $\alpha =  5 \, ^\circ$ to $\alpha =  7 \, ^\circ$ \cite{PhdKoehler} resulting in higher target densities.\\
Next to the cluster beam density, the average number $N$ of atoms per cluster is an important key figure of clusters. It is predictable via the empirical Hagena's scaling law \cite{Hagena_2}

\begin{equation}
N = A_N \left( \frac{\gamma^*}{1000} \right)^{\gamma_N},
\end{equation}

wherein $A_N = 33$ and $\gamma_N = 2.35$ are empirical values for a Hagena parameter $\gamma^* > 1800$ \cite{Hagena_1, Hagena_2}. The Hagena parameter $\gamma^*$ 
	
\begin{equation}
\gamma^* = \frac{\hat{k} p_0 \left( \frac{0.74 d_{\text{n}}}{\tan \alpha_{1/2}} \right)^{0.85}} {T_0^{2.29}}
\end{equation}

predicts the formation of clusters and a value of $\gamma^* > 1000$ leads to an immense formation of clusters. Here, $\hat{k}_{\text{H}_2} = 184$ is a gas dependent constant \cite{k}, $d_{\text{n}}$ the narrowest inner nozzle diameter, and $\alpha_{1/2} = 3.5 \, ^\circ $ the expansion half angle of the nozzle. \\
The Hagena parameter and scaling law are based on measurements performed with nitrogen, carbon dioxide, and argon. Therefore, the equations hold true for these gas types. In case of hydrogen, mass distribution measurements were successfully performed at the University of M{\"u}nster and resulted in a good agreement within an additional constant underestimation by a factor of $2.6$ \cite{PhdKoehler}. Moreover, it was shown that the cluster mass distribution can be well described by a log-normal distribution
\begin{equation}
f \left( x, \mu, \sigma \right) = \frac{1}{x \sigma \sqrt{2 \pi}} \exp \left( - \frac{\left( \ln \left( x \right) - \mu \right)^2}{2 \sigma^2 } \right),
\end{equation}
with the scaling parameters $\mu$ and $\sigma$, and the arithmetic mean
\begin{equation}
N = \exp \left( \mu + \frac{\sigma^2}{2} \right).
\end{equation}
The diameter of a cluster $d_{\text{Cluster}}$ can be calculated using the average number $N$ of atoms per cluster 

\begin{equation}
d_{\text{Cluster}} = 2 \left( \frac{3 m N}{4 \pi \rho_{\text{Cluster}}}	\right)^\frac{1}{3}.
\end{equation}

The density of a cluster $\rho_{\text{Cluster}}	$ is assumed to be the solid density of the used gas and $m$ is the mass of the hydrogen atom. The estimated average cluster diameters for the measurements presented within this publication are shown and compared to measurements using Mie-scattering in Table \ref{tab:Cluster}. \\

\section{Cluster size measurements using Mie-scattering}
\begin{figure}
\includegraphics[scale=0.40]{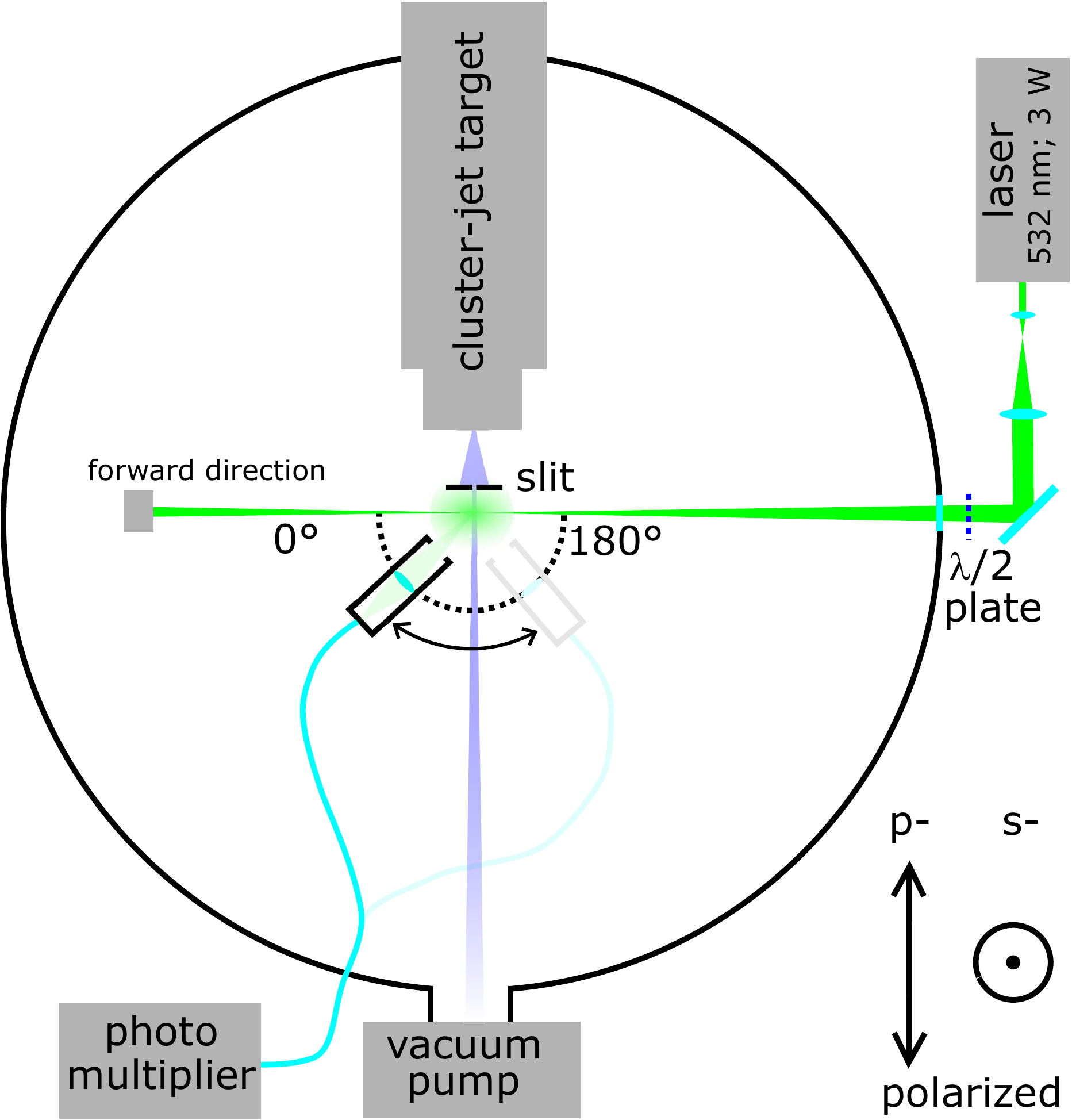}
\caption{\label{Mie_setup} Setup of the performed Mie-scattering measurement. The laser beam is focused into the cluster-jet and the scattered light is collected at different angles with respect to the laser forward direction for different polarizations S and P.}
\end{figure}

\begin{figure}[htb]
\includegraphics[scale=0.30]{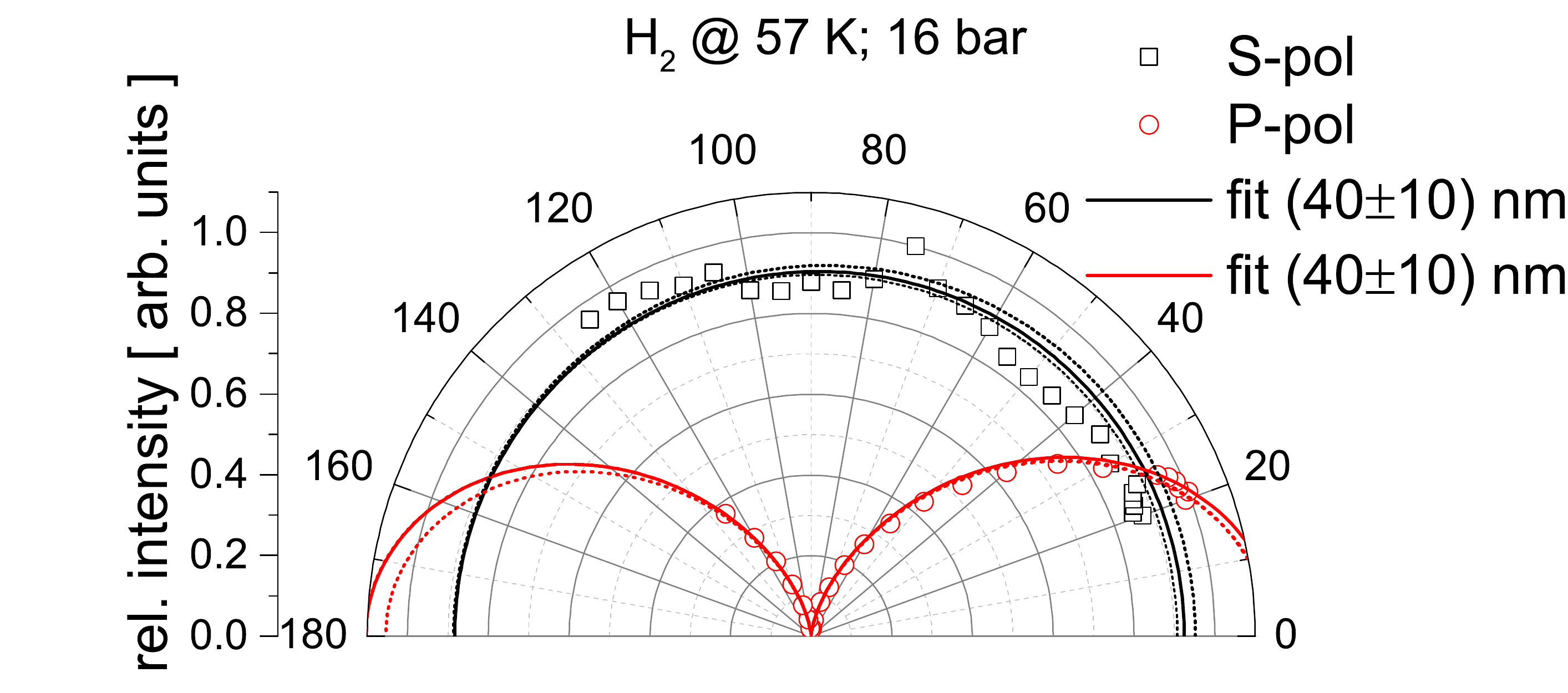}
\includegraphics[scale=0.30]{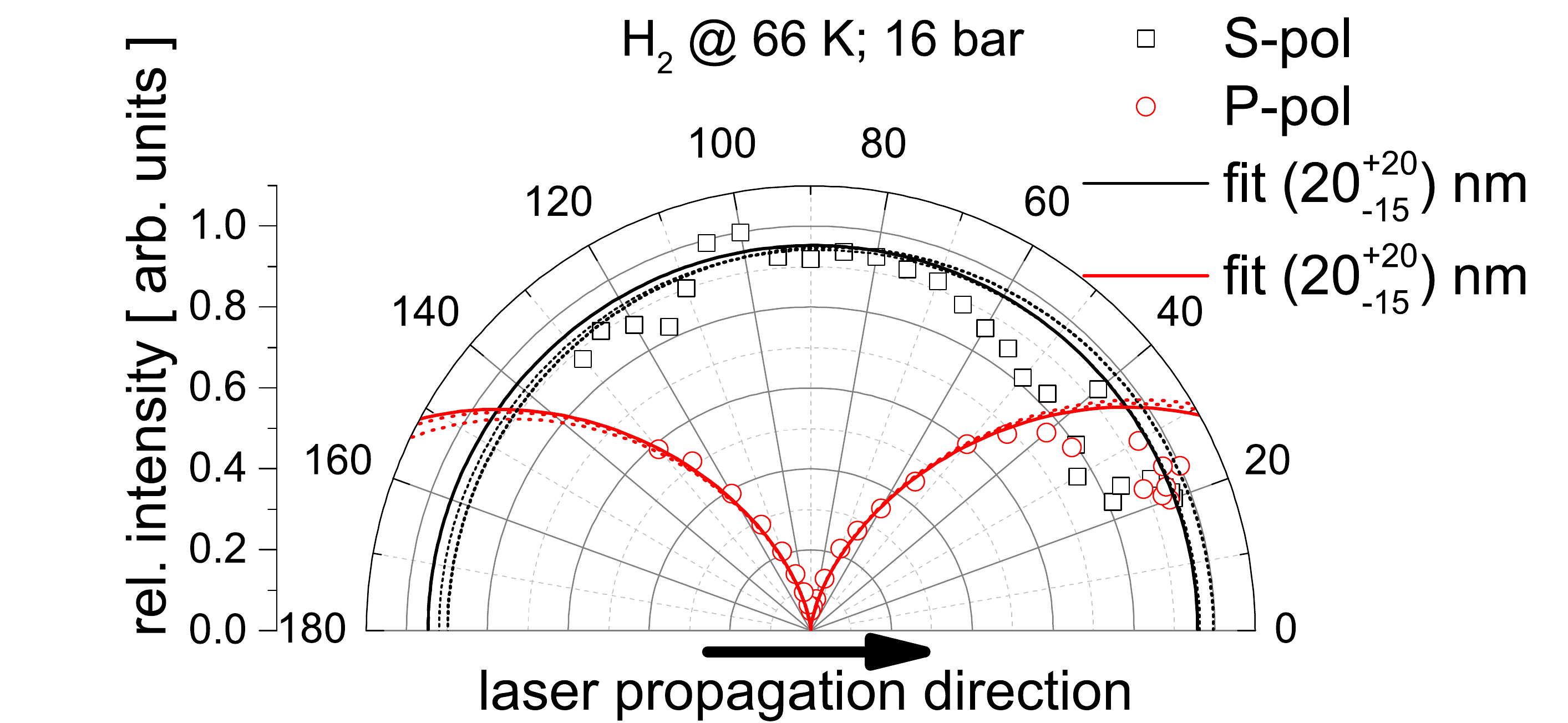}
\caption{\label{mie_measure} Measurement of the different cluster size diameters: H$_{\mathrm 2}$ at 57\,K: $d$=(40$\pm$10)\,nm and H$_{\mathrm 2}$ at 66\,K: $d$=(20$^{+20}_{-15}$)\,nm.}
\end{figure}

For further studies of the underlying proton-acceleration process and to compare with theoretical predictions described in section III an experimental verification of the cluster size was carried out. \textit{Mie-scattering} \cite{Mie08} can be used for an accurate measurement of the cluster size. As illustrated in Fig. \ref{Mie_setup} a 3\,W frequency-doubled solid state laser at 532\,nm was used, focused to a spot size of about 50\,$\mu$m onto the cluster-jet 28\,mm downstream from the Laval nozzle. In the following, the laser forward direction is defined as 0$^{\circ}$. Before entering the vacuum a $\lambda$/2 plate was installed to change the laser polarization direction. The cluster-jet was limited by a slit with a width of about 5\,mm in the laser propagation direction to define a clear interaction volume. The scattered light was collected by an achromatic lens with a focal length of 75\,mm and an acceptance angle of 23\,mrad imaging the interaction point onto a glass fibre. This imaging line was placed on a rotatable stage and could be rotated between 20\,$^{\circ}$ and 130\,$^{\circ}$ in the laser incidence plane. The fibre was coupled onto a photomultiplier (PM - Hamamatsu R928) outside the vacuum chamber and the relative voltage was recorded as signal.
The angularly dependent scattering of the perpendicular- (S) or parallel- (P) polarized laser light, with respect to the detection plane, was measured. The intensity $I_{\mathrm 0}\left(\Theta, \alpha(N) \right)$ scattered by a single cluster, composed of $N$ atoms, into an angle $\Theta$, is given by \cite{Boh83}
\begin{equation}
I_{\mathrm 0}\left(\Theta, \alpha(N) \right)=\left(\frac{\lambda}{2\pi R}\right)^2 I_{\mathrm 0} i_{\mathrm \perp;\parallel}(\Theta, \alpha(N) ),
\end{equation}
where $\alpha(N)=\frac{2\pi a(N)}{\lambda}$ is the size parameter, which depends on the incident wavelength $\lambda$ and the Mie-radius $a$. $R$ is the distance between the detector and the interaction point, $I_{\mathrm 0}$ is the initial intensity and $i_{\mathrm \perp;\parallel} \equiv i_{\mathrm \perp;\parallel}/\alpha^3$ is a fitting parameter for the S- and P-polarized case, respectively. A change of the polarization of the incident laser light results in a different scattering behaviour. While perpendicular polarized light is scattered homogeneously, only having an asymmetric intensity in forward and backward direction, the parallel polarized light has an additional minimum at 90$^{\circ}$ with respect to the direction of the incident light.\\
Since there can be more than one cluster in the scattering volume, the total intensity scattered is given by the integral over all clusters consisting of $N$ atoms in the volume
\begin{equation}
I(\Theta)=V_{\mathrm 0} \int_{\mathrm 0}^{\infty}I_{\mathrm 0}\left(\Theta, \alpha(N) \right) f(N) \, \text{d}N.
\end{equation}
Here, $f(N)$ is the particle size distribution depending on the one hand on the spatial position within the cluster jet and on the other hand defined by the log-normal distribution of the statistical generation process of the clusters. The application of a 5\,mm slit width directly behind the cluster nozzle leads to a decrease of the spatial distribution of the cluster-jet beam and by this to a well defined scattering volume. However, although principally possible, due to the limited covered scattering angle range and angular resolution, in this special case this method does not allow to resolve the log-normal particle size distribution. Based on that, the assumption was made, that the measured values are a reflection of the average size of the clusters. Following the theoretical approach described by \textit{C. F. Bohren} \cite{Boh83}, the theoretical Mie-scattering distributions for P- and S-polarized light were calculated and fitted to the measured values. Note that in the presented case the use of different polarizations instead of different wavelengths results \cite{Pra12} in a verification of the measurements. For the calculated values of $i_{\mathrm \perp;\parallel}$ the refractive index depending on the laser wavelength $\lambda$ needs to be taken into account. For hydrogen $n_{\text{H}_2; \text{ liq.}}$=1.117 \cite{Joh37} was assumed.
Fig. \ref{mie_measure} illustrates the measured values (symbols) and the theoretical fitted diameters (lines). The dashed lines represent the corresponding error range. A smaller scattering source results in a more spatially isotropic scatter behaviour. For this reason, the errors become large in this case. The results are summarized in Table \ref{tab:Cluster}. The predicted cluster diameters using the Hagena formula and the measured cluster diameters via the Mie-scattering setup are in good agreement within the uncertainties.

\section{Results of laser-driven acceleration using the hydrogen cluster-jet target}
\begin{figure}
\includegraphics[scale=0.25]{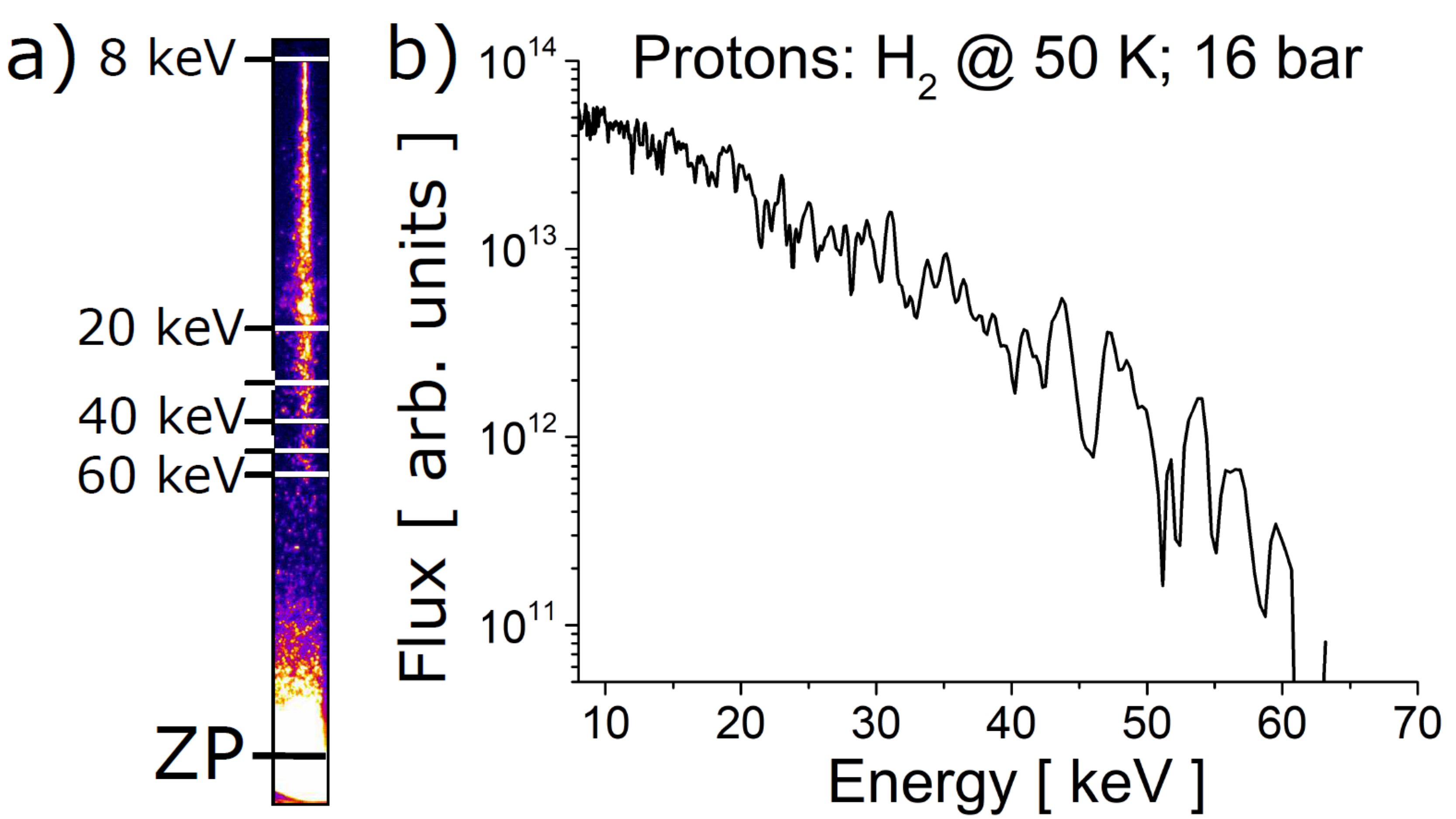}
\caption{\label{protonspectrum} a) Raw data image of a Thompson-parabola spectrometer. b) Evaluated energy spectrum with a cut-off energy of roughly 60\,keV.}
\end{figure}

Preliminary measurements of proton acceleration using the hydrogen cluster-jet target at the ARCTURUS laser facility of the University of D\"usseldorf were performed. A 200\,TW Ti:Sapphire laser with an energy of 7\,J  before compression, a pulse duration of 30\,fs and with a beam diameter of 80\,mm was used to irradiate the cluster-jet beam. The laser pulse was focused by an f/2 off-axis parabola to a focal diameter of 4\,$\mu$m, corresponding to an intensity of 2$\times$10$^{20}$\,W/cm$^2$. For the results reported here, the hydrogen gas in front of the Laval nozzle had a temperature of roughly 50\,K and a pressure of 16\,bar resulting in a cluster size of roughly $d$=45\,nm.\\ 
The proton spectrum was measured along the laser forward direction by means of a Thompson-parabola spectrometer \cite{Gwy04}. Figure \ref{protonspectrum}a shows the raw image of proton track on the detector. In Fig. \ref{protonspectrum}b the energy spectrum is presented which was extracted employing a MATHLAB routine. In this measurement, maximum energies of roughly 60\,keV were observed. Moreover, a high shot-to-shot stability of the proton energy spectrum was demonstrated by $50$ subsequent shots and will be published elsewhere.\\
The calculated cluster target density in combination with the average number of atoms within one cluster predicted via the Hagena law allows for the estimation of the number of clusters within the focal volume (V$_{\text{Foc}}$=189\,$\mu$m$^3$), defined as the volume within one Rayleigh-length of the laser focus. In the case of the proton-acceleration measurement the mean cluster number in the focal volume was determined to be $13$. Consequently there is more than one single cluster located in the interaction volume which forms the basis for stable and reproducible proton spectra. These and other further results on the physics of the acceleration process, which is Coulomb-explosively driven, will be presented in detail in another publication.\\

\begin{table*}
\caption{\label{tab:Cluster}Calculated and measured target parameters and characteristics for the parameter sets investigated. The density was estimated based on the calculated gas flow through the nozzle.}
\begin{ruledtabular}
\begin{tabular}{c|c|c|c|c|c}
 gas & pressure / bar & temperature / K & density / (atoms/cm$^3$) & predicted cluster & measured cluster \\
   &  &  &  &  diameter / nm &  diameter / nm \\

	 \hline
 	hydrogen & $16.0 \pm 0.1$ & $57.0 \pm 0.1$ & $\left( 3.4 \pm 0.2 \right)  \times 10^{16}$ & $35.3  \pm 1.1 $& $40 \pm 10$ \\  
					 & $16.0 \pm 0.1$ & $66.0 \pm 0.1$ & $\left( 2.9 \pm 0.2 \right)  \times 10^{16}$ & $27.1 \pm 0.8  $& $20 ^{+20}_{-15}$ \\  
\end{tabular}
\end{ruledtabular}
\end{table*}

\section{Conclusion}
A cluster-jet target was designed, built and successfully taken into operation at the University of M\"unster. This target can be used with different gases and allows for a stable and continuous flow of pure target material. Additionally, this cluster-jet target was implemented at the ACTURUS laser facility of the University of D\"usseldorf for laser-driven proton acceleration. A detailed technical description of the cluster source demonstrates the features and possibilities of such a target. Theoretical calculations of the essential cluster parameters, like density and cluster size using the Hagena relation for different stagnation conditions were performed. Moreover, the cluster sizes were measured using a Mie-scattering technique and the results demonstrate a good agreement with the Hagena prediction within the uncertainties. First successfully performed measurements using hydrogen clusters irradiated by a high-intensity 5\,Hz laser prove the operational reliability of the target system by observing protons in the multi-keV range. Based on these results, a newly designed cluster-jet target is currently under construction at the University of M\"unster. This target will significantly extend the region of accessible stagnation conditions even towards the generation of hydrogen clusters from the liquid phase. This will result in significantly larger cluster sizes and higher densities to achieve considerable higher kinetic energies of the accelerated protons.

\section{Acknowledgement}
We acknowledge financial support by the German Federal Ministry of Education and Research (BMBF) joint research project No. 05K16PM3 and No. 05K16PFA. The work provided by the teams of the mechanical and electronic workshops at M\"unster University is very much appreciated and we thank them for the excellent manufacturing of the various components. We thank A. Alejo from Queens University Belfast for the spectrometer evaluation routine.\\

\section{References}

\bibliography{references}

\end{document}